\documentclass{aa}  
\usepackage{xcolor}
\usepackage{graphicx}
\usepackage{txfonts}
\usepackage[citecolor=cyan]{hyperref}
\begin{document} 

    \title{Unveiling very young O stars}
   \subtitle{Two Galactic O2V((f*))z in Westerlund 2}
   \author{A. Roman-Lopes
          \inst{1}
          }

   \institute{Department of Astronomy, Universidad de La Serena,
              Av. Raul Bitran \#1302, La Serena, Chile\\
              \email{aroman@userena.cl}
             }

   \date{Received ; accepted }

 
  \abstract
   {O-type stars are known to significantly contribute to both the dynamics and evolution of galaxies. Massive and luminous, they probably control and regulate the galaxies star formation rates. The stellar feedback generated by such cosmic beasts can strongly affect the local star formation rate, with effects in the current (and future) generations of low and intermediate mass stars, and possibly also in the disruption process of the galaxies' giant gas reservoirs.}  
   {For this work I performed a redetermination of the spectral types and effective temperatures of the Galactic O-type stars MSP182, MSP183, MSP199, VPHAS-01338, and VPHAS-01273.}
   {From a careful examination of the spectral features present in the blue optical spectral region, it was possible to identify several nitrogen lines usually only seen in the blue optical spectra of O2-O3 stars. From the nitrogen ionic equivalent width ratios measured in the spectra of MSP182, MSP183, MSP199, VPHAS-01338, and VPHAS-01273, and in those of standard stars of the O2-O4 spectral types, earlier spectral types and hotter effective temperature values were derived.}
   {{Two O2V((f*))z, together with three new O3 V stars are now firmly identified in the Westerlund 2 region.  Besides RFS1 in NGC3603, the O2 stars found in Westerlund 2 are the only other exemplars known to date in the Milky Way. From the nitrogen equivalent width line ratios measured in the spectra of standard stars of the O2-O4 spectral types, linear relations  between the N\textsc{iv}$\lambda$4058/N\textsc{iii}$\lambda$4640 ratio and the effective temperature in the 47000K-51000K range were derived. Based on my spectroscopic analysis of the science targets and the use of a HRD, a mean heliocentric distance of 5kpc to Westerlund 2 was computed, a result that is in line with the mean heliocentric distance of 5.3$\pm${1.5} kpc obtained from the associated \textit{Gaia} DR3 parallaxes and distances.}}
  {{The Westerlund 2 massive stars studied in this work probably share a common evolutionary process that might be representative of the evolutionary ages of a large fraction of the cluster's O-type stellar population, which seems to be much younger than 1 Myrs.}}
%

   \keywords{Stars: early-type, massive
               }

   \maketitle
%

\section{Introduction}

\paragraph{}O-type stars are known for their elevated temperature, substantial size, and high luminosities, playing a pivotal role in shaping the evolution and dynamics of galaxies. Their lifespans, which are relatively brief on astronomical timescales (typically lasting a few million years), offer valuable insights into the temporal progression of galactic populations. Functioning as stellar powerhouses, O-type stars emit substantial energy in the form of intense ultraviolet radiation, stellar winds, and eventual supernova events. Consequently, they exert a pronounced influence on the interstellar medium (ISM), significantly contributing to the dynamics and evolutionary processes within galaxies. In this sense, they probably regulate the star formation rates within galaxies since the stellar feedback generated by such cosmic beasts can affect both the formation of subsequent star generations and the dispersion process of giant molecular clouds \citep{ferr00,smith10,dale11,hopkins12}.

The O2 is the earliest spectral type among O stars. It was introduced by \citet{walb02}, designating early O stars with distinct characteristics in their spectral lines. Their spectra show very weak (or absent) He\textsc{i} and {N\textsc{iii} 4640} lines, which -- combined with a higher excitation manifested through prominent {N\textsc{iv} 4058\AA~ and N\textsc{v} 4602-4621\AA} transition lines -- sets O2 stars apart from their later type siblings. The unique spectral fingerprint among normal O-type stars suggests that the O2 type boasts the highest effective temperatures among dwarf, giant, and super-giant stars.
In the Galaxy, only a couple of O2 super-giant stars are known to date: HD93129A in Trumpler 14 and RFS3 in the periphery of NGC 3603, with both being classified as O2 If* \citep{maiz14} and \citep{roman16}. Surprisingly, no O2 III stars are known in the Milky Way to date, and the only O2 dwarf is RFS1 \citep{roman16} which is found at about 1 arcmin south of the center of the super-star cluster NGC3603.
As can be noticed, the observed paucity of exemplars of the O2 type underscores the rarity of such stars that formed from metal-rich molecular gas clouds, as normally found in the Galaxy.
On the other hand, in the lower metallicity environment of the Magellanic Clouds, there are at least ten known O2 stars: three O2 If* (Brey77, RMC136a5, and cl* NGC 2070 MH 59), one O2 IIIf* (VFTS16), and six dwarfs (Walborn 2, BI237, BI253, VFTS169, VFTS468, and VFTS512), with many of them spread around the 30 Doradus starburst region \citep{river12} and \citep{walb14}. The higher number of O2 stars found there when compared to the Galaxy can be understood if we consider that, in the presence of a lower metallicity gas environment, the most massive stars produced in the associated molecular clouds tend to be hotter for a given spectral type \citep{mok07,mass09,walb14,mart17}.
For this work I performed a reexamination of the spectral types and effective temperatures of the Galactic O-type stars MSP182, MSP183, MSP199, VPHAS-01338, and VPHAS-01273, communicating the identification of two O2 and three O3 main sequence stars in the Westerlund 2 cluster, which were previously misclassified as being of later spectral types.

\section{New O2 and O3 main sequence stars in Westerlund 2}

The Galactic O-type stars MSP182, MSP183, MSP199, VPHAS-01338, and VPHAS-01273 (hereafter the O-star sample) are found in the direction of the Westerlund 2 stellar cluster. They were recently studied by \citet{drew18} using non-local thermodynamic equilibrium model-atmosphere fitting techniques applied to the hydrogen and helium line profiles present in the associated blue optical spectra. The authors concluded that the earliest source of the sample would be MSP183, a O3 V star that was previously classified as such by \citet{rauw11}. In the case of the remaining sources of the O-star sample, MSP182, MSP199, VPHAS-01338, and VPHAS-01273, the authors concluded that all are O4 V stars. However, from a careful examination of the spectral features present in the 4000$\AA$-5000$\AA$ wavelength range of the blue optical spectra of the sample (which are not presented in their entirety in the original article), it is seen that there are (in all of them) absorption and emission nitrogen lines in a combination that is usually only present in the spectra of stars of the O2-O3 types. This is of the utmost importance as it is well known that the canonical method of using the He\textsc{i}/He\textsc{ii} line ratio \citep{conti71} and \citep{mat89} might fail in the determination of the effective temperature (or spectral types) of the hottest O-type stars. Instead, \citet{walb02} and \citet{river12b} suggested complementing the spectral analyses using specific nitrogen ionic lines such as the N\textsc{iii} 4634-4640-4642, N\textsc{iv} 4058, N\textsc{v} 4602, and/or N\textsc{v} 4621 transitions present in the associated blue optical spectral range.

\subsection{Spectral types of the O-star sample based on blue optical spectra of very early-O standard stars}

In order to reestimate the spectral types of the O-star sample sources, in this section their XSHOOTER blue optical spectra are compared with those of the standard stars HD269810 - O2 III((f*)), BI237 - O2V((f*))z, BI253 - O2 V((f*)), HD64568 - O3 V((f*))z, and HD46223 - O4 V((f)) \citep{walb02,sota11,river12,walb14}, taken from the ESO phase-III archive\footnote{http://archive.eso.org/cms.html}. Table 1 lists the instrument name and the ESO program ID from which the associated spectral data were obtained. All spectra used in the comparison were normalized using the SPLOT task of IRAF, with all of them matching the same spectral resolution R=6655 that corresponds to the smallest spectral resolution of the spectra (in the 4000$\AA$-5000$\AA$ wavelength range) used in this work. Finally, in the case of spectral data for which the original resolution is superior to that of the O-star sample, the template spectrum was degraded to match the spectral resolution R=6655. 
The results are shown in Figure 1. The  spectra of the O-star sample are shown in black, while those from the O-type standards are seem in gold and red. Also, the most important spectral features are indicated by the vertical black labels. Next, I describe and comment on the results obtained for each source of the O-star sample in detail.

   \begin{figure*}
   \centering
   \includegraphics[width=17cm]{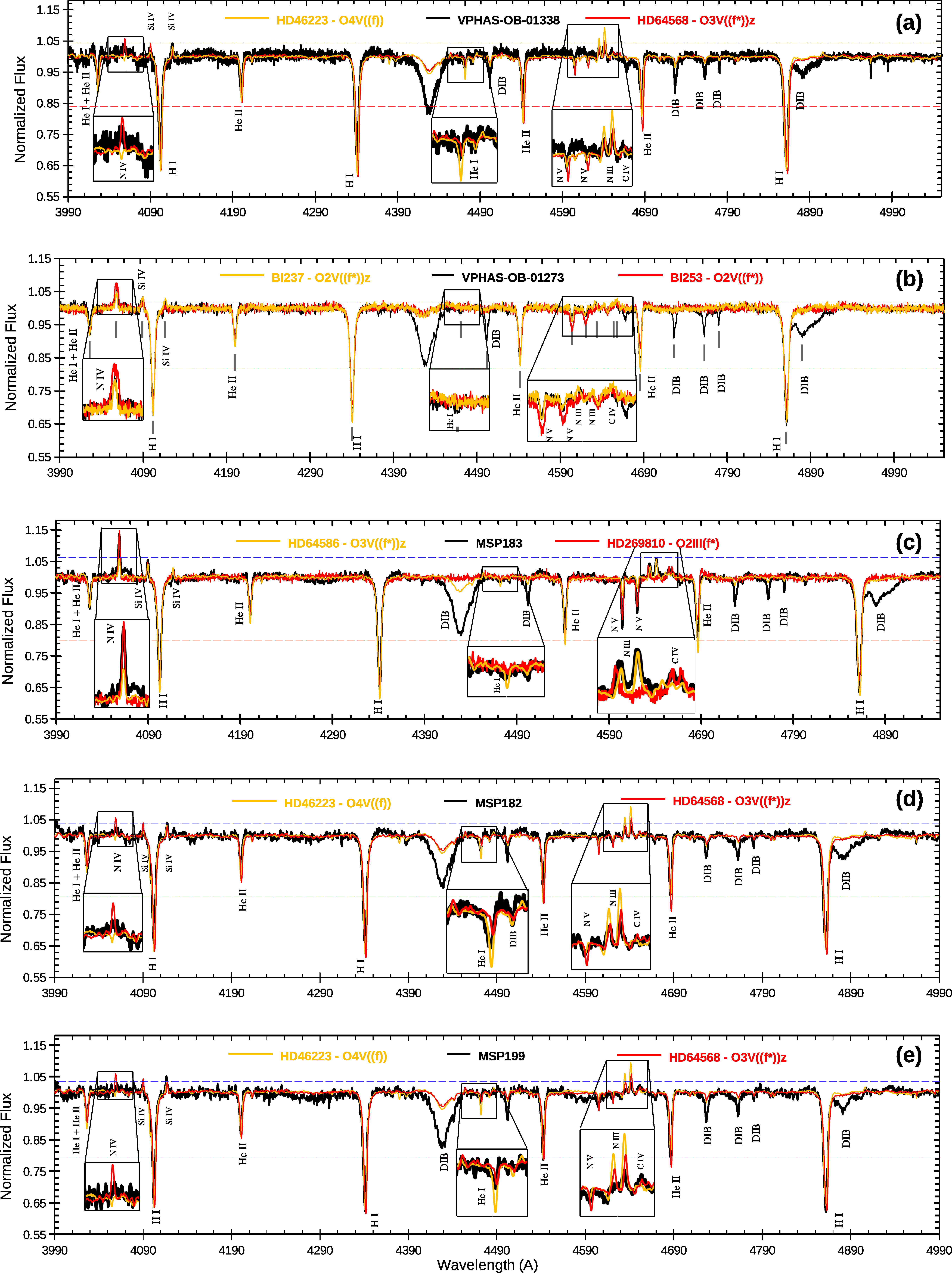}

   \caption{Blue optical spectra for the O-type star sample. Rectified  spectra in the 3990$\AA$-4990$\AA$ wavelength range of the Galactic O-type stars VPHAS-01338, VPHAS-01273 MSP183, MSP182, and MSP199, together with those of the standard stars HD269810 - O2 III((f*)), BI237 - O2 V((f*))z, BI253 - O2 V((f*)), HD64568 - O3 V((f*))z, and HD46223 - O4 V((f)), taken from the ESO phase-III archive. In the case of VPHAS-01273 and MSP183, their 4686$\AA$ lines correspond to the most intense He absorption line in the 4000$\AA$-5000$\AA$ wavelength range (with their minimum intensities being indicated by the horizontal dashed, light red lines), which is a behavior characteristic of the Vz class \citep{walb14}, so they are very young O2 stars. The maximum intensity of the N\textsc{iii} 4640 lines are indicated by the horizontal dashed, gray lines.}
              \label{FigGam}%
    \end{figure*}
%


\subsection{VPHAS-01338: O3 V((f*))}

{The blue optical spectrum of VPHAS-01338 is shown in Figure 1(a) together with the spectra of the Galactic stars HD64568 (O3 V((f*))z) and HD46223 (O4 V((f))). It can be observed that its He\textsc{i} 4471$\AA$ absorption line is much weaker than that of HD46223, a behavior that indicates that  VPHAS-01338 is hotter. Its blue optical spectrum resembles that of HD64568 well, the Galactic standard of the class, with the NIV 4058$\AA$ emission line showing a similar intensity to the N III 4640$\AA$ one \citep{walb02}. Taking into account the observed similarity between their He I, N III, N V, and C IV lines (see the insets shown in Figure 1(a), the spectral type O3 V((f*)) is assigned for VPHAS-01338.}

\subsection{VPHAS-01273: O2 V((f*))z}
  
{In Figure 1 the rectified blue optical  spectrum of the VPHAS-01273 source (black) is shown, together with those of BI237 (gold) and BI253 (red), which are both standards of the O2 V((f*)) class \citep{walb02,river12,walb14}. On the one hand, the former shows very intense He\textsc{ii} absorption lines at 4200$\AA$, 4542$\AA,$ and 4686$\AA$, and the absence of any of the He\textsc{i} lines normally seen in the optical spectra of most O-type stars indicate that the star's spectral type is certainly earlier than O4 \citep{walb02}. Its He\textsc{ii} absorption line at 4686$\AA$ is seen in pure absorption (the horizontal dashed line indicates the minimum of its observed intensity), so a dwarf class is assigned. On the other hand, the presence of the N\textsc{iv} 4058$\AA$ line in emission, together with absence of He I lines indicate that this object probably belongs to the hottest lineage of O-type stars. Last but not the least, the spectral feature seen at 4686$\AA$ corresponds to the {most intense He absorption line} in the 4000$\AA$-5000$\AA$ wavelength range, which is a behavior characteristic of the Vz class \citep{walb14}, so VPHAS-01273 is likely very young. From the comparison of the VPHAS-01273 spectrum with those of the BI237 and BI253, one can see that the three are very similar, which suggests that VPHAS-01273 is a Galactic exemplar of the O2 V((f*))z type.}

\subsection{MSP183: O2 V((f*))z}

{The blue optical spectrum of MSP183 shown in Figure 1(c) presents very intense He\textsc{ii} 4200$\AA$, 4542$\AA,$ and 4686$\AA$ absorption lines. The absence of any of the He\textsc{i} lines that are usually seen in the optical spectra of most O stars indicates that MSP183's spectral type is possibly earlier than O3 \citep{walb02}.
The comparison of its blue optical spectrum with that of the O3 V((f*))z star HD64568 (the Galactic standard of the type) shows that it presents a stronger He\textsc{i} 4471$\AA$ absorption line, which, together with a C IV 4658 line much weaker than the one in the science spectrogram \citep{walb02}, leaves no doubt that MSP183 is a hotter star. 
On the other hand, the comparison of its spectrum with that of the O2 star HD269810 shows a rather similar behavior with one of the few noticeable differences between them being related to their He\textsc{ii} 4686$\AA$ absorption lines. In the case of HD269810, its He\textsc{ii} line is seen partially (weakly) in emission (so luminosity class III); whereas, in the MSP183 optical spectrogram, it is observed in pure absorption (luminosity class  V). As in the case of VPHAS-01273, its He\textsc{ii} 4686$\AA$ absorption line is the most intense He transition in the 4000$\AA$-5000$\AA$ wavelength range, and the Vz class is secured \citep{walb14}. Finally, the NIV 4058$\AA$ emission line in the MSP183 data is much stronger than NIII 4640$\AA$, so the O2 type is the best choice in that case \citep{walb02}}.

   \begin{figure}
   \centering
   \includegraphics[width=9.0cm]{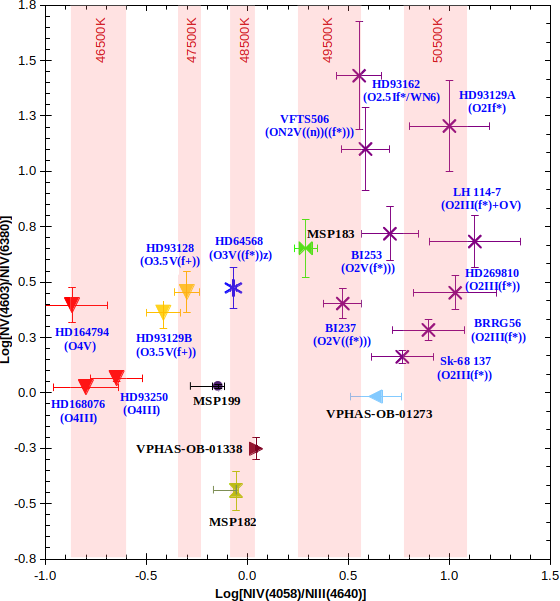}

   \caption{Log[N V$\lambda$4603/N\textsc{iv}$\lambda$6380] $\times$ Log[N\textsc{iv}$\lambda$4058/N\textsc{iii}$\lambda$4640] diagram for the O2 (violet crosses), O3 (blue star), O3.5 (gold triangles), and O4 (red triangles) spectroscopic standard stars (labeled by the blue text). The two O2V((f*)) stars of Westerlund 2, MSP183, and VPHAS-01273, and the three O3 V$\lambda$(f*)) stars, VPHAS-01338, MSP182, and MSP199, are also indicated by the black text.}
              \label{FigGam}%
    \end{figure}

\subsection{MSP182 (O3.5 V(f+)) and MSP199 (O3 V((f*))}

{The two remaining sources of the O-star sample are MSP182 and MSP199. Their blue optical spectra are shown in Figure 1(d) and 1(e), together with those of HD64568 (O3 V((f*))z) and HD46223 (O4 V((f))). The He\textsc{i} 4471$\AA$ absorption lines are seen much weaker in the MSP182 and MSP199 spectrograms; although, in the case of MSP182, they are stronger than that seen in the O3 standard spectrum. Also, and as in the case of VPHAS-01338, the N\textsc{iii} 4640$\AA$ emission lines appear much weaker than that seen in the HD46223 spectrogram. In the case of MSP182, its N\textsc{iv} 4058$\AA$ emission line appears weaker than that of N\textsc{iii} 4640$\AA$, which together enables its spectral types to be estimated to be O3.5 \citep{walb02}. On the other hand, in the case of MSP199, its N\textsc{iv} 4058$\AA$ emission line appears with a similar intensity to that of N\textsc{iii} 4640$\AA$ (see the horizontal dashed line positioned in the peak of the N\textsc{iii} 4640$\AA$ emission line), so O3 type is assigned in its case. Finally, their He\textsc{ii} 4686$\AA$ lines are both seen in pure absorption, and the dwarf classes are assigned to MSP182 and MSP199.}
In what follows, I further explore the optical spectra of the O-star sample by using the information provided by the N\textsc{iii}$\lambda$4640, N\textsc{iv}$\lambda$4058, N\textsc{iv}$\lambda$6380, and N\textsc{v}$\lambda$4603 ionic lines.

\section{Spectral types and effective temperatures from nitrogen ionic equivalent width ratios}

The canonical method of using the He\textsc{i}/He\textsc{ii} line ratios \citep{conti71,mat89} fails in the determination of the effective temperatures for the earliest O stars, with a degeneracy of possible solutions in the associated parameters' space. In order to brake this degeneracy, \citet{walb02} proposed that the scheme in which ionic transitions of the nitrogen atom, namely the ratio N\textsc{v}$\lambda$4058/N\textsc{iii}$\lambda$4640, be used instead. Later, \citet{river12b} provided theoretical support to it by modeling the observed nitrogen ion line ratios. They concluded that {once the luminosity class is known}, the effective temperatures of O2-O4 stars can be estimated well from the N\textsc{v}$\lambda$4058/N\textsc{iv}$\lambda$4640 line ratios.

\subsection{Spectral types}

\begin{table*}
\caption{List of ESO spectra used in this work, in which the star ID, spectral type (from the literature), instrument, and ESO programs are indicated.}             
\centering                          
\begin{tabular}{c c c c }        
\hline\hline                 
ID & SpType (literature) & Instrument & ESO-Program \\    
\hline                        
   HD93129A & O2 If* & FEROS & 079.D-0564(A) \\      
   HD93162 & O2.5 If*/WN6 & FEROS & 0102.A-9010(A) \\
   LH 114-7 & O2 III(f*)+O V & XSHOOTER & 106.211Z \\
   HD269810 & O2 III(f*) & UVES+XSHOOTER & 70.D-0164(A)+106.211Z \\
   BRRG56 & O2 III(f*) & UVES+XSHOOTER & 074.D-0109(A)+084.D-0142(A) \\ 
   Sk-68 137 & O2 III(f*) & XSHOOTER & 106.211Z \\ 
   BI237 & O2 V((f*)) & UVES+XSHOOTER & 074.D-0109(A)+106.211Z \\ 
   BI253 & O2 V((f*)) & UVES+XSHOOTER & 074.D-0109(A)+106.211Z \\
   VFTS506 & ON2 V((n))((f*)) & XSHOOTER & 106.211Z \\
   HD64568 & O3 V((f*))z & FEROS & 096.A-9039(A) \\
   BRRG48 & O3 V((f*))z & XSHOOTER & 106.211Z \\
   HD93128 & O3.5 V(f+) & FEROS & 086.D-0997(A) \\
   HD93129B & O3.5 V(f+) & FEROS & 089.D-0975(A) \\
   HD93250 & O4 III & FEROS & 073.D-0609(A) \\
   HD168076 & O4 III & FEROS & 077.D-0146(A) \\ 
   HD46223 & O4 V((f)) & FEROS & 086.D-0997 \\
   MSP183 & O3 V((f)) & XSHOOTER & 095.D-0843(A) \\ 
   VPHAS-OB1-1338 & O4 V & XSHOOTER & 095.D-0843(A) \\ 
   VPHAS-OB1-1273 & O4 V & XSHOOTER & 095.D-0843(A) \\ 

\hline                                   
\end{tabular}
\end{table*}

\begin{table}
\caption{O-star sample: (1) spectral types and effective temperatures from \citet{drew18}; (and 2) spectral types and effective temperatures from this work.}             
\centering                          
\begin{tabular}{l l l l l}        
\hline\hline                 
Star & SpType\textsuperscript{1} & Teff\textsuperscript{1} & SpType\textsuperscript{2} & Teff\textsuperscript{2}
\\
  &  & (K) &  & (K)\\
\hline                        
VPHAS-1273 & O4 V & 45200 & O2V((f*))z & 49700 \\
VPHAS-1338 & O4 V & 46300 & O3 V((f*)) & 48500 \\
MSP182 & O4 V & 45500 & O3 V((f*)) & 48300 \\
MSP183 & O3 V((f)) & 49000 & O2V((f*))z & 49000 \\
MSP199 & O4 V & 49100 & O3.5 V((f)) & 48150 \\

\hline                                   
\end{tabular}
\end{table}

It is possible to provide new estimates for the spectral types of the O-star sample sources by comparing their observed N\textsc{iv}$\lambda$4058/N\textsc{iii}$\lambda$4640 and N\textsc{v}$\lambda$4603/N\textsc{iv}$\lambda$6380] equivalent width (EW) ratios, with those from known spectroscopic standards of the O2, O3, and O4 types. In Table 1 the list of the ESO Phase 3\footnote{http://archive.eso.org/cms.html} dataset containing the optical spectra of O-type standards used in this work is shown.
All spectra used in my measurements were carefully continuum rectified and degraded to the same spectral resolution as previously explained in Section 2.

   \begin{figure}
   \centering
   \includegraphics[width=9.0cm]{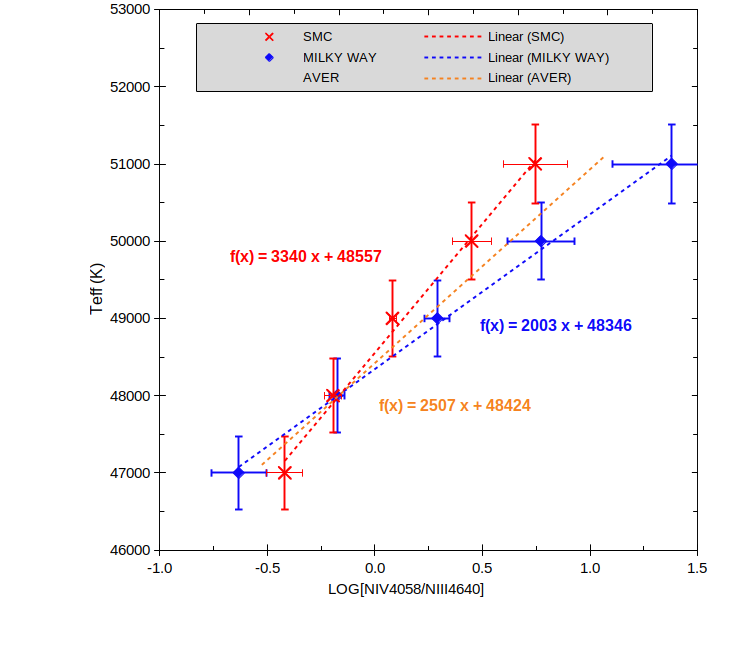}

   \caption{ Log[N\textsc{iv}$\lambda$4058/N\textsc{iii}$\lambda$4640] versus effective temperature diagram. The theoretical relations are represented, respectively, by the blue (Milky Way) and red (SMC) symbols. It can be seen that in both cases, there is a linear relation between the effective temperature and the Log[N\textsc{iv}$\lambda$4058/N\textsc{iii}$\lambda$4640]. The linear trend appears steeper in the case of the SMC models, so, as consequence, any O2 star should be hotter than its siblings in the Galaxy. One can also see that in the mentioned temperature range, the maximum difference between the effective temperature values of SMC and Milky Way stars (occurring close to the value Log[N\textsc{iv}$\lambda$4058/N\textsc{iii}$\lambda$4640] = 1.0) is only $\eqsim$ 1000K. On the other hand, it can be observed that for a O3 star (Log[N\textsc{iv}$\lambda$4058/N\textsc{iii}$\lambda$4640]=0), both Milky Way and SMC stars should have similar effective temperatures (see text).}
              \label{FigGam}%
    \end{figure}

In Figure 2 the Log[N\textsc{v}$\lambda$4603/N\textsc{iv}$\lambda$6380] $\times$ Log[N\textsc{iv}$\lambda$4058/N\textsc{iii}$\lambda$4640] diagram is presented in which the empirical positions of O2 (violet crosses), O3 (blue star), O3.5 (gold triangles), and O4 (red triangles) spectroscopic standard stars are shown. They are labeled by the bold blue text. It can be seen that the distinct spectral types appear clearly separated from each other, mainly in the Log[N\textsc{iv}$\lambda$4058/N\textsc{iii}$\lambda$4640] dimension of the diagram, with the O2 stars' group being clearly displaced from the later types by about 0.5 dex. The two new O2V((f*))z stars VPHAS-01273 and MSP183 are found just in this part of the Log[N\textsc{v}$\lambda$4603/N\textsc{iv}$\lambda$6380] $\times$ Log[N\textsc{iv}$\lambda$4058/N\textsc{iii}$\lambda$4640] diagram.
Also, accordingly to \citet{walb02}, the O3 spectral type should correspond to EW[N\textsc{iv}$\lambda$4058]/EW[N\textsc{iii}$\lambda$4640] line ratio = 1 (e.g., Log[N\textsc{iv}$\lambda$4058/N\textsc{iii}$\lambda$4640] = 0). From Figure 2, it can be seen that besides HD64568 (the Galactic standard of the O3 class), sources VPHAS-01338, MSP182, and MSP199 all have Log[N\textsc{iv}$\lambda$4058/N\textsc{iii}$\lambda$4640] ratio values close to zero, which suggests that they are very likely also O3 stars \citep{walb02}. These results are in line with those obtained in Section 2, in which the O3 V spectral type was assigned for VPHAS-01338, MSP182, and MSP199.

\subsection{Effective temperatures}

As mentioned above, \citet{river12b} performed a theoretical study of the behavior of the nitrogen ion line ratios in the case of O2-O4 stars. Particularly
in their Figure 3, they present the theoretical relations between the EW of N\textsc{iii}$\lambda$4640 and N\textsc{iv}$\lambda$4058 lines, as a function of the effective temperature of O-type stars in the Galaxy (Milky Way), and in the Small Magellanic Cloud (SMC), for models with gravity and nitrogen abundance values of log g = 4.0 and [N] = 7.78 dex, respectively.

From the work of \citet{river12b}, I proceeded by tabulating the N\textsc{iii}$\lambda$4640 and N\textsc{iv}$\lambda$4058 EW values for the Milky Way and SMC models considering the {thin wind case} for constant temperature values in the interval 47000K-51000K. In Figure 3, the Log[N\textsc{iv}$\lambda$4058/N\textsc{iii}$\lambda$4640] versus effective temperature diagram obtained from the computed ratio values is shown. There, the theoretical relations are represented, respectively, by the blue (Milky Way) and red (SMC) symbols. It can be seen that in both cases, there is a linear relation between the effective temperature and the Log[N\textsc{iv}$\lambda$4058/N\textsc{iii}$\lambda$4640]. The linear trend appears steeper in the case of the SMC models, so, as consequence, any O2 star there should be hotter than its sibling in the Galaxy. However, one can also see that in the mentioned temperature range, the maximum difference between the effective temperature values of SMC and Milky Way stars (occurring close to the value Log[N\textsc{iv}$\lambda$4058/N\textsc{iii}$\lambda$4640] = 1.0) is only $\eqsim$ 1000K. On the other hand, it can be observed that for a O3 star (Log[N\textsc{iv}$\lambda$4058/N\textsc{iii}$\lambda$4640]=0), both Milky Way and SMC stars should have similar effective temperatures. This occurs because the use of EW {ratios} minimizes the effects of metallicity differences between stars in the Milky Way and in the SMC. This is a very fortunate behavior as it occurs just in the temperature range in which the canonical method of using the He\textsc{i}/He\textsc{ii} line ratios \citep{conti71,mat89} starts to fail. Linear fittings to the theoretical points were performed in each case, and the results are shown in Figure 3 by the dashed red and blue lines, together with the corresponding linear relations derived from the linear fits. They can be used to estimate the effective temperatures of the earliest O-type stars, in the ranges -0.75$<$Log[N\textsc{iv}$\lambda$4058/N\textsc{iii}$\lambda$4640]$<$1.5 in the case of the Galaxy and in the range -0.5$<$Log[N\textsc{iv}$\lambda$4058/N\textsc{iii}$\lambda$4640]$<$0.75 in the case of SMC early O-type stars. As a complement, the linear relation in the intermediate case between the Milky Way and SMC is also provided. It is represented in Figure 3 by the dashed gold line which corresponds to the linear fit applied to the average of the Log[N\textsc{iv}$\lambda$4058/N\textsc{iii}$\lambda$4640] values.
   \begin{figure}
   \centering
   \includegraphics[width=8cm]{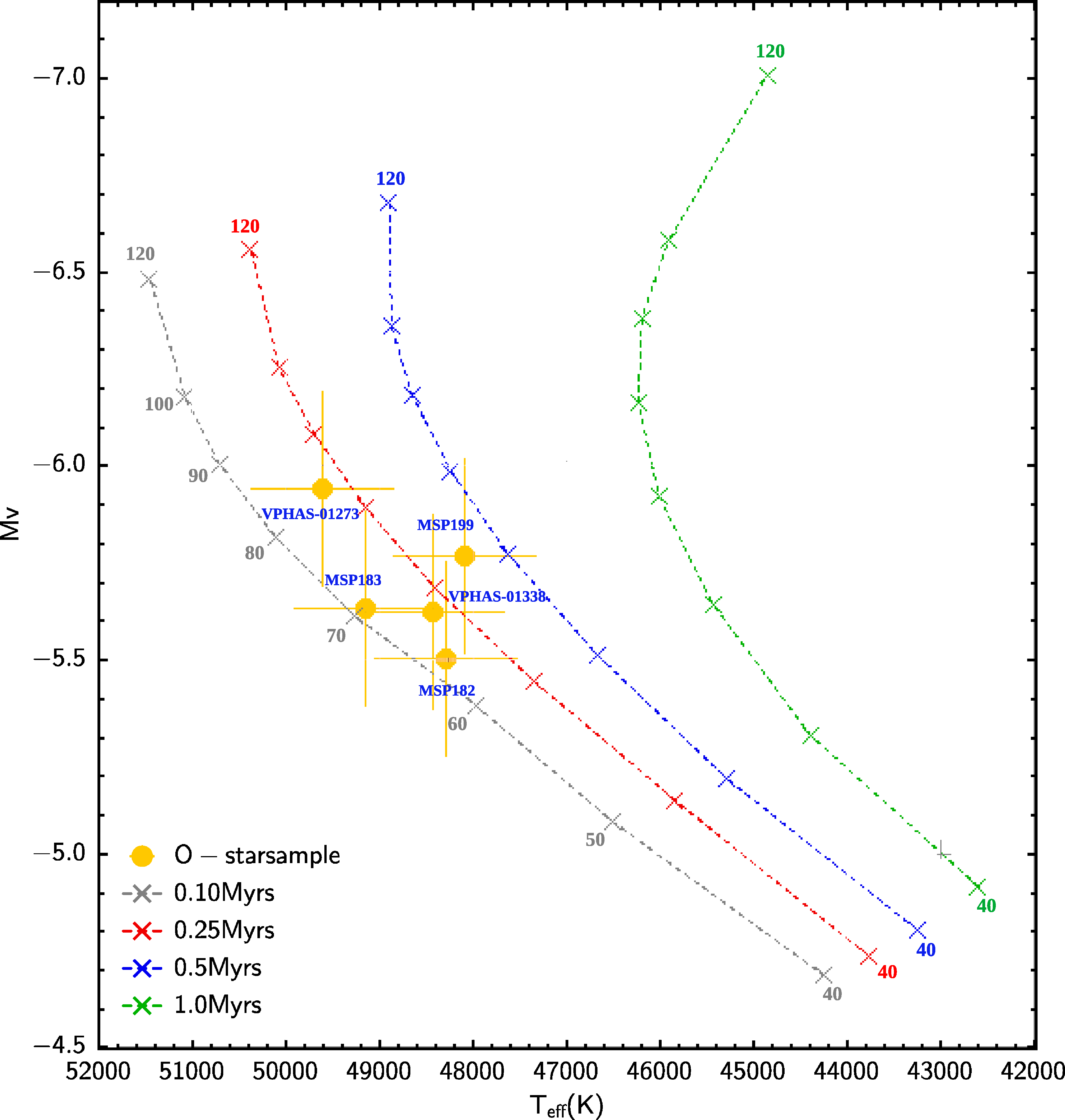}

   \caption{HRD of the stars studied in this work for a heliocentric distance of 5kpc. The MSP and VPHAS sources are represented by the golden-filled circles, while the 0.10, 0.25, 0.5, and 1.0 Myrs PARSEC isochrones (40M$_{\odot}$-120M$_{\odot}$ mass range) are shown by the dashed gray, red, blue, and green lines, respectively. Based on their position in this diagram, they are all very young massive stars, probably much younger than 1Myrs. The observed positions of the VPHAS sources \#1273 and \#1338 indicate that they are very massive main sequence stars with theoretical initial masses in the range 60M$_{\odot}$-80M$_{\odot}$, similar to those of sources MSP182, MSP183, and MSP199 that are placed in the Westerlund 2 cluster center (for more on this subject, see the main text).}
              \label{FigGam}%
    \end{figure}
One can use the derived linear relations to estimate approximate values for the effective temperatures of the standard O-type stars shown in the Log[N\textsc{v}$\lambda$4603/N\textsc{iv}$\lambda$6380] $\times$ Log[N\textsc{iv}$\lambda$4058/N\textsc{iii}$\lambda$4640] diagram of Figure 2. I used the corresponding linear equations shown in Figure 3 to compute the average SMC and Milky Way effective temperature values for O2, O3, O3.5, and O4 stars. They are represented in the plot by the vertical light red lines. The different widths of the lines indicate the approximate uncertainty ranges for the temperatures 46500K (O4), 47500K (O3.5), 48500K (O3), and 49500K-50500K (O2). Interestingly, the three standards of the O2V class present Log[N\textsc{iv}$\lambda$4058/N\textsc{iii}$\lambda$4640] values around $\eqsim $0.6 dex, while the O2 III class members present higher values around $\eqsim0.9$ dex, which is equivalent to an increase in the effective temperature of about 1000K. This behavior is probably related to the fact that for stars with stronger stellar winds, the N\textsc{iv}$\lambda$4058/N\textsc{iii}$\lambda$4640 ratio could be less reliable as a tool for the determination of effective temperatures of giants and super-giants of the O2 spectral type since the derived values  would  probably be overestimated by at least $\eqsim$ 1000K.

\subsection{The Hertzsprung-Russel diagram of the O-star sample}

   \begin{figure}
   \centering
   \includegraphics[width=9.0cm]{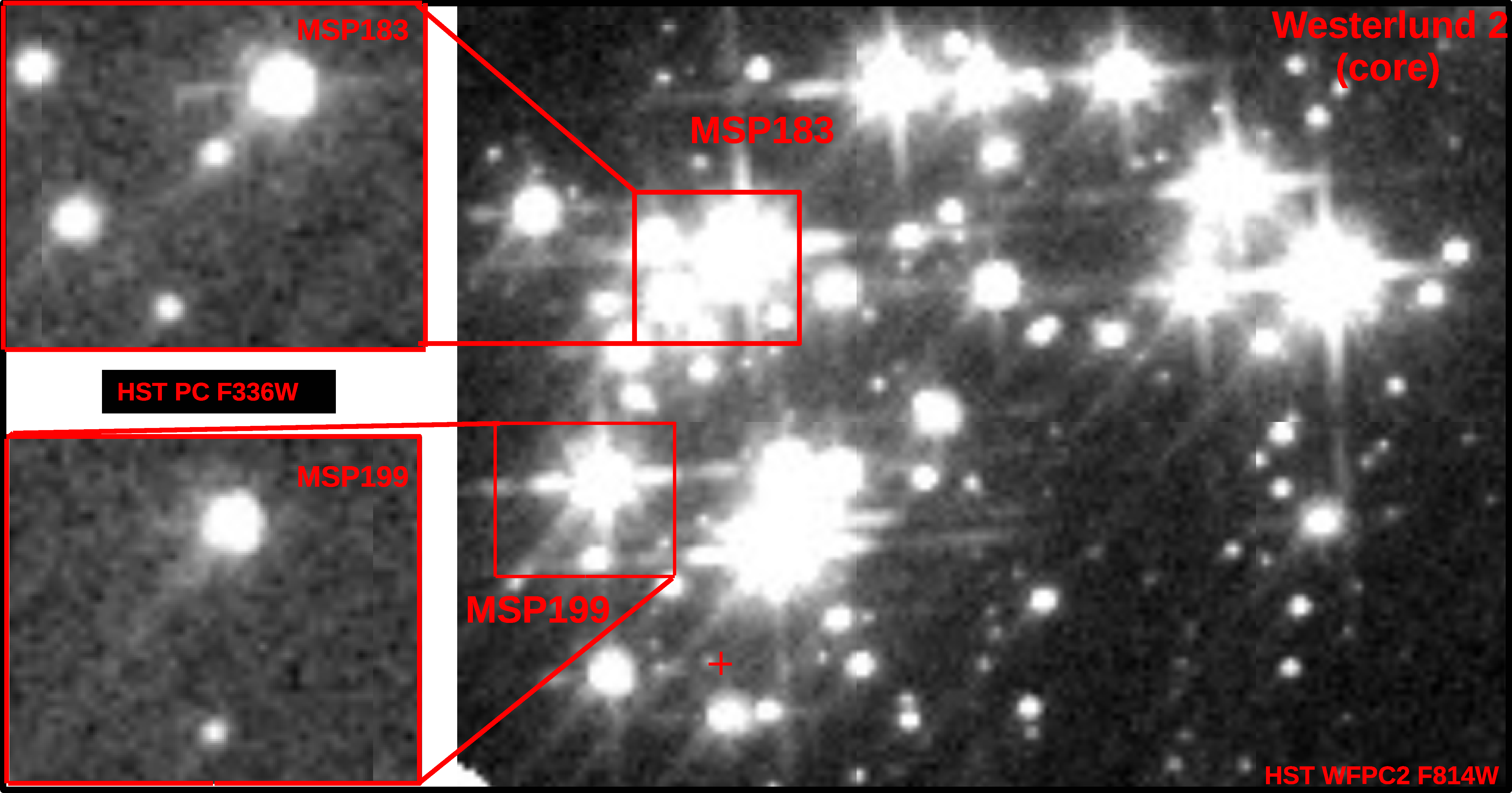}

   \caption{Detailed view of the central part of the Westerlund 2 cluster. Right: HST WFPC2 F814W image of the central part of the Westerlund 2 cluster. The higher resolution HST PC F336W cut-off image of the region around MSP183 (shown to the left) reveals that it appears blended with several fainter sources in the WFPC2 F814W image as well as in the lower resolution images taken from ground-based telescopes. This provides an explanation as to why it was found to be so luminous (and massive) in previous studies of Westerlund 2 \citep{rauw07,drew18}.}
              \label{FigGam}%
    \end{figure}

{The Hertzsprung-Russel diagram (HRD) of the O-star sample is shown in Figure 4, with the stars represented by the golden-filled circles. Their effective temperatures (shown in Table 2) correspond to the values calculated using the Galactic linear relation derived in Section 3, applied to the observed N\textsc{iv}$\lambda$4058 and N\textsc{iii}$\lambda$4640 EW values presented in Table 3.
Their absolute magnitudes M$_V$ were computed using the V-band magnitudes taken from the work of \citet{lasker08} (VPHAS-1273 and VPHAS-1338), and the V-band magnitudes of MSP183 and MSP199 were computed from the F555W and F814W magnitudes provided by the Hubble Space Telescope Legacy Archive\footnote{https://hla.stsci.edu/hlaview.html}, using the transformation equations derived by \citet{harris18}. I chose to use this procedure in the case of these two stars because, from an inspection of the HST PC F336W image of the central part of the Westerlund 2 cluster, I realized that MSP183 appears blended (see Figure 5) in the HST WFPC2 F814W lower spacial resolution image (and probably also in those taken from ground-based telescopes). This fact provides a further explanation of why MSP183 appears so luminous (and massive) in previous studies of Westerlund 2 \citep{rauw07,drew18}.
On the other hand, similar to MSP199, MSP182 is also seen relatively isolated in the HST archive images of Westerlund 2, and in its case, I chose to use the V-band magnitude value derived by \citet{rauw07}. The visual absorption {values of the sources of the O-star sample were} corrected using the A$_V$ values computed by \citet{drew18}. In the case of MSP183, its computed A$_V$ value was probably affected by its {fainter companions}. Fortunately, as it was placed only at a few arc-seconds from MSP199 (see Figure 5), it is reasonable to assume that its visual absorption is probably similar, so I used A$_V=6.4$ mag instead. 
The positions of the 0.10 Myrs, 0.25 Myrs, 0.5 Myrs, and 1.0 Myrs {PAdova and TRieste Stellar Evolution Code (PARSEC))\footnote{http://stev.oapd.inaf.it/cgi-bin/cmd} isochrones \citep{bressan12}} (40M$_{\odot}$-120M$_{\odot}$ mass range) are indicated by the dashed gray, red, blue, and green lines. Regarding the uncertainties on the HRD parameters, in the case of the effective temperatures, it is shown in Section 3.2 that the uncertainties in the Teff for values in the range 48000K-50000K are probably on the order of less than $\eqsim$ 500K. However, in order to be conservative, 1500K was assumed to be the total uncertainty instead. On the other hand, in the case of absolute visual magnitudes, an uncertainty of 0.25 magnitudes was assumed. It can be seen that the five stars present (virtually) the same visual absolute magnitudes, with an observed dispersion on M$_V \eqsim$ 0.5 mag. Regarding the mean heliocentric distance used in the computations, it was set considering that the observed position of the stars in the HRD must be consistent with the most recent results dictated by the massive star evolution theory.\ This, in the case of dwarf stars of the O2 and O3 types, corresponds to the mass range 60M$_\odot$-70M$_\odot$ \citep{mok07,mart17}, which in turn sets a mean heliocentric distance of $\eqsim$5kpc. \\
Although the heliocentric distance of Westerlund 2 is an important problem, it is well beyond the main goal of this work. However, it is useful and necessary to provide some comments on it. The heliocentric distance of 5kpc used in the construction of the HRD is the natural consequence of the consistency between the overall results presented here, and the need to satisfy the range of masses dictated by the current theory of massive star formation, through the use of state-of-the-art isochrones. As result, the 5 kpc value is the most probable mean heliocentric distance to the five stars of the O-star sample. On the other hand, it is possible to check the consistency of my result, by comparing it with the heliocentric distances provided by \textit{Gaia} DR3.
In order to do that, the heliocentric distances for the five stars of the O-star sample were taken from the \textit{Gaia} EDR3 catalog of \citet{bailer21}, and the mean values for rgeo and rpgeo were computed. The results of the calculations are 5293$\pm{1523}$ pc and 5257$\pm{1555}$ pc, respectively, with the quoted errors corresponding to the standard deviation values of the mean.}

\section{Discussion}

The O2 stars VPHAS-01273 and MSP183, together with RFS1 \citep{roman16} are the only other dwarfs of the type identified to date in the Milky Way. They are placed in Westerlund 2, which, with the NGC 3603 starburst cluster, are the only known sites harboring Galactic dwarfs of the O2 type. 
As very young massive stars, MSP183, together with its siblings MSP182 and MSP199, are found exactly at the place predicted by star formation theory: in the center of their parental cluster. On the other hand, the odd novelty comes from the fact that VPHAS-01273 and VPHAS-01338 are not found in a stellar cluster. They are indeed found {isolated} in the field, at 11.6\arcmin and 21.2\arcmin from the closest known massive star cluster, respectively. How {60M$_\odot$-80M$_\odot$ stars} can be found so far from their likely parental cluster is unclear. A further study on this theme is beyond the context of this Letter; however, next, I briefly comment on some aspects (and consequences) of the subject.

\subsection{The earliest runway dwarf stars known in the Galaxy}

\citet{drew18} proposed an explanation for the current location of VPHAS-01273 and VPHAS-01338, based on their study of radial velocities and proper motions of candidate runaway stars placed around the Westerlund 2 cluster. The authors postulated that, in the case of VPHAS-01273 and VPHAS-01338, the stars traveled from the central parts of the nearest massive star cluster, which in their case is Westerlund 2. {The authors} concluded that in both cases, their observed space velocities meet the currently accepted standard runaway criterion, calculating that the stars were possibly ejected from Westerlund 2 at $\eqsim$ 500,000 and 800,000 years ago,  respectively. Massive star formation theory predicts that stars such as VPHAS-1273 and VPHAS-01338 should evolve very quickly, leaving the main sequence in no more than a few million years. Considering the estimated travel times, it is reasonable to assume that if the stars were expelled from their parental cluster, it is very likely that it occurred in the very beginning of their main sequence lifetimes. Under such conditions, some of them should present some sort of spectral signature evidencing their very early evolutionary stage. In fact, such spectral evidence does exist, at least in the case of VPHAS-1273. Its He\textsc{ii} 4686$\AA$ line is the {most intense He} {absorption} line in the 4000$\AA$-5000$\AA$ wavelength range and, as previously mentioned, a behavior characteristic of stars of a very young age \citep{walb14}. From an astrophysical perspective, it is desirable to check the consistency of a very small age for the stars, by comparing their positions in the HRD relative to those of the PARSEC theoretical isochrones presented in Figure 4. {There they are seen as very young stars with evolutionary ages in the 100,000-500,000 year range.}
From this reasoning, it is possible to conclude that VPHAS-1273 and VPHAS-01338 are probably the earliest runway O-dwarf stars known in the Galaxy.

\subsection{Consequences for the age of the massive stellar population of Westerlund 2}

From an astrophysical perspective, it is useful to return to the HRD of Figure 4. The correct understanding of the physical processes that impact (or potentially govern) the temporal evolution of very massive Galactic clusters such as NGC3603 and Westerlund 2 can be partially deepened if one adequately considers the results obtained here for the O-star sample as a whole.
We shall start by looking at the positions of the stars in the HRD, not only relative to the evolutionary isochrones, but also how the ages of the stars compare to each other. On the one hand, VPHAS-01273 and MSP183 are quite young with ages inferred from the theoretical isochrones in the range of {100,000-500,000 years}. On the other hand, it turns out that MSP182, MSP199, and VPHAS-01338 appear to not only show very similar absolute visual magnitudes, but very similar evolutionary ages, a youth condition that is in line with the Vz behavior observed in both O2 stars (see section 2). In fact, the entire O-star sample appears in the HRD as a very cohesive group of sources, which, from an astrophysical point of view, indicates that they probably share a common evolutionary process, perhaps representative of the entire {massive stellar population} of the cluster. This could suggest that the Westerlund 2 {very massive stellar population} is probably as young (or even younger) than the less massive members of the cluster, which have estimated mean ages in the range $\eqsim$0.5Myrs-2.0Myrs \citep{carr13}, \citep{zeid15,zeid18}.

\section{Concluding remarks}

For this work I performed a redetermination of the spectral types and effective temperatures of the Galactic O-type stars MSP182, MSP183, MSP199, VPHAS-01338, and VPHAS-01273. The main results are summarized as follows:

   \begin{enumerate}
      \item Two O2V((f*))z (MSP183 and VPHAS-01273), together with three new O3-type stars (MSP182, MSP199, and VPHAS-01338) are spectroscopically confirmed in Westerlund 2.  Besides RFS1 in NGC3603, the two O2 stars found in Westerlund 2 are the only other exemplars known to date in the Milky Way. The O2 star MSP183 is found to be very close to several fainter sources in the HST PC F336W image of the central part of Westerlund 2, providing an explanation on why it was found so luminous (and massive) in previous studies of Westerlund 2 \citep{rauw07}, \citep{drew18}.
      \item From the nitrogen EW line ratios measured in the spectra of standard stars of the O2-O4 spectral types, linear relations  between the N\textsc{iv}$\lambda$4058/N\textsc{iii}$\lambda$4640 ratio and the effective temperature in the 47000K-51000K range were derived. As expected, the linear trend appears steeper in the case of the low metallicity models. However, the {maximum} difference between the effective temperature of SMC and Milky Way O2 stars, which occurs close to the edge of the mentioned temperature range, corresponds to only $\eqsim$ 1000K. {On the other hand, in the case of O3 stars, both Milky Way and SMC stars should have similar effective temperatures. This occurs because the use of EW {ratios} probably minimizes the effects of metallicity differences between stars in the Milky Way and in the SMC. This is a very fortunate behavior as it occurs just in the temperature range in which the canonical method of using the He\textsc{i}/He\textsc{ii} line ratios  starts to fail.}
      \item {Based on the results obtained from my spectroscopic analysis of the science targets and the use of a HRD, a mean heliocentric distance of 5kpc to Westerlund 2 was computed, a result that is in line with the mean heliocentric distance of 5.3$\pm${1.5} kpc obtained from the associated \textit{Gaia} DR3 parallaxes and distances.}
      \item The Westerlund 2 massive stars studied in this work probably share a common evolutionary process that might be representative of the evolutionary ages of a large fraction of the cluster's O-type stellar population, which seems to be much younger than 1 Myrs.
   \end{enumerate}

\begin{acknowledgements}
      ARL thanks the anonymous referees for the constructive reports that contribute to improve the interpretation of the results obtained in this work.
      This work is based (in part) on data obtained from the ESO Science Archive Facility with DOI(s): https://doi.org/10.18727/archive/71, https://doi.org/10.18727/archive/50, https://doi.org/10.18727/archive/24. It is also based [in part] on observations made with the NASA/ESA Hubble Space Telescope, and obtained from the Hubble Legacy Archive, which is a collaboration between the Space Telescope Science Institute (STScI/NASA), the Space Telescope European Coordinating Facility (ST-ECF/ESA) and the Canadian Astronomy Data Centre (CADC/NRC/CSA). ARL acknowledges the developers of the TOPCAT software (http://www.starlink.ac.uk/topcat/), as well as the PARSEC team (ttp://stev.oapd.inaf.it/cmd) for providing the state of the art isochones models used in this work.
\end{acknowledgements}

\begin{table}
\caption{EW values measured from the N\textsc{iv}$\lambda$4058 and N\textsc{iii}$\lambda$4640 lines present in the XSHOOTER optical spectra of the sources of the O-star sample. V-band magnitudes of the sources of the O-star sample are provided: MSP182 magnitudes are from \citet{rauw07}; MSP183 and MSP199 - F555W and F814W magnitudes are provided by the Hubble Space Telescope Legacy Archive transformed using the results of \citet{harris18}; and the VPHAS sources are from \citet{lasker08}. The A$_V$ values are from the work of \citet{drew18}.}             
\centering                          
\begin{tabular}{c c c c c}        
\hline\hline                 
Star ID & EW[N\textsc{iv}$\lambda$4058] & EW[N\textsc{v}$\lambda$4640]& V & Av\\
  & (\AA) & (\AA)& (mag) & (mag) \\
\hline                        
VPHAS-1273 & -0.17(2) & -0.04(1) & 14.22 & 6.67 \\
VPHAS-1338 & -0.11(1) & -0.10(1)& 13.87 & 6.17 \\
MSP182 & -0.07(1) & -0.08(1)& 14.49 & 6.40 \\
MSP183 & -0.31(1) & -0.16(1)& 14.53 & 6.37 \\
MSP199 & -0.05(1) & -0.07(1)& 13.86 & 6.40 \\

\hline                                   
\end{tabular}
\end{table}


\begin{thebibliography}{}  

  \bibitem[Bailer-Jones et al.(2021)]{bailer21} Bailer-Jones, C. A. L., Rybizki, J., Fouesneau, M., Demleitner, M., Andrae, R. 2021, AJ, 161, 147
  \bibitem[Bressan et al.(2012)]{bressan12} Bressan, Alessandro, Marigo, Paola, Girardi, Léo., Salasnich., Bernardo, Dal Cero, Claudia, Rubele, Stefano, Nanni, Ambra 2012, MNRAS, 427, 127 
  \bibitem[Carraro et al.(2013)]{carr13} G. Carraro, G., Turner, D., Majaess, D., Baume, G. 2013, A\&A, 550, 50
  \bibitem[Conti \& Alschuler(1971)]{conti71} Conti P. S., Alschuler W. R., 1971, ApJ, 170, 325   
  \bibitem[Dale \& Bonnell(2011)]{dale11} Dale, James E., Bonnell, Ian 2011, MNRAS, 414, 1, 321
  \bibitem[Drew et al.(2018)]{drew18} Drew, J. E., Herrero, A., Mohr-Smith, M., Monguió, M., Wright, N. J., Kupfer, T., Napiwotzki, R. 2018, MNRAS, 480, 2109
  \bibitem[Ferrara \& Tolstoy(2000)]{ferr00} Ferrara, Andrea, Tolstoy, Eline 2000, MNRAS, 313, 2, 291
  \bibitem[Harris(2018)]{harris18} Harris, W. E. 2018, AJ, 156, 296 
  \bibitem[Hopkins et al.(2012)]{hopkins12} Hopkins, Philip F., Quataert, Eliot, Murray, Norman 2012, MNRAS, 421, 4, 3522
  \bibitem[Lasker et al.(2008)]{lasker08} Lasker, Barry M., Lattanzi, Mario G., McLean, Brian J., Bucciarelli, Beatrice, Drimmel, Ronald, Garcia, Jorge, Greene, Gretchen, Guglielmetti, Fabrizia, Hanley, Christopher, Hawkins, George, Laidler, Victoria G., Loomis, Charles, Meakes, Michael, Mignani, Roberto, Morbidelli, Roberto, Morrison, Jane, Pannunzio, Renato, Rosenberg, Amy, Sarasso, Maria, Smart, Richard L., Spagna, Alessandro, Sturch, Conrad R., Volpicelli, Antonio, White, Richard L., Wolfe, David, Zacchei, Andrea 2008, AJ, 136, 735  
  \bibitem[Maiz-Apellaniz et al.(2014)]{maiz14} 2014, ApJS, 211, 10
  \bibitem[Martins \& Palacios(2017)]{mart17} F. Martins, A. Palacios 2017, A\&A, 598, 56
  \bibitem[Massey et al.(2009)]{mass09} Massey, Philip, Zangari, Amanda M., Morrell, Nidia I., Puls, Joachim, DeGioia-Eastwood, Kathleen, Bresolin, Fabio, Kudritzki, Rolf-Peter 2009, ApJ, 692, 618
  \bibitem[Mathys(1989)]{mat89} Mathys G., 1989, A\&AS, 81, 237
  \bibitem[Mokiem et al.(2007)]{mok07} M. R., Mokiem, A. de Koter, C. J. Evans, J. Puls, S. J. Smartt, P. A. Crowther, A. Herrero, N. Langer, D. J. Lennon, F. Najarro, M. R. Villamariz, and J. S. Vink 2007, A\&A, 465, 1003
  \bibitem[Rauw et al.(2007)]{rauw07} Rauw, G., Manfroid, J., Gosset, E., Nazé, Y., Sana, H., De Becker, M., Foellmi, C., Moffat, A. F. J. 2007, A\&A, 463, 981
  \bibitem[Rauw, Sana \& Naz\'e(2011)]{rauw11} Rauw, G., Sana, H., Naz\'e, Y. 2011, A\&A, 535, 40
  \bibitem[Rivero-Gonzalez et al.(2012)]{river12} Rivero Gonz\'alez, J. G., Puls, J., Najarro, F., Brott, I. 2012, A\&A, 537, 79
  \bibitem[Rivero-Gonzalez et al.(2012)]{river12b} Rivero Gonz\'alez, J. G., Puls, J.,  P. Massey, Najarro, F. 2012, A\&A, 543, 95
  \bibitem[Roman-Lopes et al.(2011)]{roman11} Roman-Lopes A., Barba, R. H., Morrell, N. I. 2011, MNRAS, 416, 501
  \bibitem[Roman-Lopes(2013)]{roman13} Roman-Lopes A. 2013, MNRAS, 435, 73
  \bibitem[Roman-Lopes et al.(2016)]{roman16} Roman-Lopes A., Franco, G.A.P., Sanmartin, D. 2016, ApJ, 823, 96  
  \bibitem[Smith et al.(2010)]{smith10} Smith, Nathan, Povich, Matthew S. , Whitney, Barbara A., Churchwell, Ed, Babler, Brian L., Meade, Marilyn R., Bally, John, Gehrz, Robert D., Robitaille, Thomas P., Stassun, Keivan G. 2010, MNRAS, 406, 2, 952
  \bibitem[Sota et al.(2011)]{sota11} Sota, A., Ma\'iz Apell\'aniz, J., Walborn, N. R., Alfaro, E. J., Barb\'a, R. H., Morrell, N. I., Gamen, R. C., Arias, J. I. 2011, ApJS, 193, 24
  \bibitem[Walborn et al.(2002)]{walb02} Walborn, Nolan R., Howarth, Ian D., Lennon, Daniel J., Massey, Philip, Oey, M. S., Moffat, Anthony F. J., Skalkowski, Gwen, Morrell, Nidia I., Drissen, Laurent, Parker, Joel Wm. 2002, AJ, 123, 2754
  \bibitem[Walborn et al.(2014)]{walb14} Walborn, Nolan R., Sana, H., Simón-Díaz, S., Maíz Apellániz, J., Taylor, W. D., Evans, C. J., Markova, N., Lennon, D. J., de Koter, A. 2014, A\&A, 564, 40
  \bibitem[Zacharias et al.(2013)]{zach13} Zacharias, N., Finch, C. T., Girard, T. M., Henden, A., Bartlett, J. L., Monet, D. G., Zacharias, M. 2013, AJ, 145, 44 
  \bibitem[Zeidler et al.(2015)]{zeid15} Peter Zeidler, p., Sabbi, E., Nota, A., Grebel1, E. K., Tosi, M., Bonanos, A. Z., Pasquali1, A., Christian, C., de Mink, S. E., Ubeda, L. 2015, AJ, 150, 78
  \bibitem[Zeidler et al.(2018)]{zeid18} Zeidler, p., Sabbi, E., Nota, A., Pasquali, A., Grebel, E. K., McLeod, A. F., Kamann, S., Tosi, 6 M., Cignoni, M., Ramsay, S. 2018, AJ, 156, 211
  
\end{thebibliography}
\end{document}